\documentstyle[preprint,revtex]{aps}

\begin{document}
\begin{flushright}
\baselineskip=15pt
UU-HEP-92/11\\
astro-ph@xxx:9212005\\
November 28, 1992\\
\end{flushright}

\begin{title}
{\bf The escape of gravitational radiation \\
from the field of massive bodies}
\end{title}

\author{Richard H. Price and Jorge Pullin}
\begin{instit}
Department of Physics, University of Utah, Salt Lake City UT 84112-1195
\end{instit}
\author{Prasun K. Kundu}
\begin{instit}
Applied Research Corporation, Landover MD 20785 and\\
Laboratory for Atmospheres, NASA/Goddard Space Flight Center,
Greenbelt, MD 20771.
\end{instit}

\begin{abstract}

We consider a compact source of gravitational waves of frequency
$\omega$, in or near a massive spherically symmetric distribution of
matter or a black hole.  Recent calculations have led to apparently
contradictory results for the influence of the massive body on the
propagation of the waves.  We show here that the results are in fact
consistent and in agreement with the  ``standard''
viewpoint in which the high frequency compact source produces the
radiation as if in a flat background, and the background curvature
affects the propagation of these waves.

\end{abstract}

\pacs{4.30.+x}

Some of the most interesting potential sources of gravitational
radiation consist of relatively compact astrophysical configurations,
in particular binary neutron stars, embedded in much larger and
massive galaxies. The standard viewpoint for radiation from such
arrangements is to separate the problem into that of the generation of
the radiation by the compact source, and that of the propagation of
the radiation through the galaxy \cite{KipRev}. The radiation generated
by the source is calculated as if the source were in a flat
background. For orbiting binary neutron stars the standard quadrupole
formalism would be a good approximation. The effect of the spacetime
curvature created by the host galaxy is then understood in terms of
its influence on the propagation of the waves to a distant observer.
In the standard viewpoint the major propagation effects are the
gravitational redshift and gravitational lensing. For a galaxy of mass
$M$ and radius $R$ these effects are of order $M/R$ and for ordinary
galaxies, and for most purposes, are very small. (We use here, and
throughout this paper, units in which $c=G=1$.)

Despite the apparent simplicity of this prevalent viewpoint, there are
some unclear issues. One of us (PK) has found mathematical relations
suggesting that the gravitational field of the galaxy might suppress,
by many orders of magnitude, the emergence of quadrupole gravitational
waves generated inside it or nearby \cite{Kundu}.  Two of us
(RP and JP) have studied the same problem and have found that the
galactic gravitational background has a minimal effect on the
emergence of the waves, and that the ``standard viewpoint'' is
valid \cite{PP1}.  It is now clear how the specific results of the two
studies can be compatible, and what the implication is for
astrophysical sources of gravitational radiation.

The  mathematics which gave rise to the appearance of
suppression was framed in the language of the Newman-Penrose \cite{NP}
(hereafter NP) formalism, and is based on the Weyl projection $\Psi_0$
in that formalism. For an outgoing solution, $\Psi_0$ takes the form
$\Psi_0 = \psi^0_0(u,\theta,\phi)r^{-5}+{\cal O}(r^{-6})$
where $u$ is retarded time.  It is well accepted that information
about outgoing gravitational waves is encoded in the shear
$\sigma=\sigma_0(u,\theta,\phi)r^{-2}+{\cal O}(r^{-4})$
and in the Bondi news function \cite{BvM} $d\sigma_0/du$.

In spherically symmetric backgrounds it is convenient to consider a
multipole decomposition and to treat separately each multipole mode.
For modes of multipole index $\ell$ we can write
$\Psi_0=\tilde{\Psi}_0\ _2Y_{\ell m} \ \ \ \psi^0_0=\tilde{\psi}^0_0 \
_2Y_{\ell m} \ \ \sigma_0=\tilde{\sigma}_0\ _2Y_{\ell m}\,$, where $\
_2Y_{\ell m}$ are the spin-weight 2 spherical harmonics.  A useful
feature of the quantities $\tilde{\Psi}_0,
\tilde{\psi}^0_0,\tilde{\sigma}_0$, with angular variables removed, is
that their real and imaginary parts correspond respectively to even-
and odd-parity modes.  If the background spacetime is a Schwarzschild
spacetime of mass $M$, and if time dependence $e^{i\omega t}$ is
assumed, the NP equations lead to the relations
\begin{equation}\label{suppor}
\tilde{\sigma_0}=\pm\frac{r^2\omega^2}{6(1\pm i\omega
M/2)}\tilde{\psi^0_0}\ ,\ \ \
\tilde{\sigma_0}=\pm
\frac{(\ell-2)!}{(\ell+2)!}\,\frac{r^2\omega^2}{\left[1\pm
\frac{(\ell-2)!}{(\ell+2)!}12i\omega M\right]}\tilde{\psi^0_0}\ ,
\end{equation}
for the quadrupole and  general $\ell$ cases.
Here the + signs apply for even-parity
perturbations, and the - signs for odd.

It is eq.~(\ref{suppor}) that suggests suppression of radiation. The
intensity of the gravitational radiation is represented by
$\tilde{\sigma_0}$. If $\tilde{\psi^0_0}$ is taken to represent the
quadrupole moment of a source, it follows that for a given quadrupole
moment oscillating at frequency $\omega$, the radiation is reduced due
to the mass of the Schwarzschild background by the factor $(1 \pm
i\omega M/2)^{-1}$, so that the radiation power flux (proportional to
$|d\sigma_0/du|^2)$ is reduced by the factor $(1+\omega^2M^2/4)^{-1}$.
For a typical galaxy $M\approx 10^{16}$ cm, and for the radiation
sources of greatest interest $\omega\approx 10^{-7} {\rm cm}^{-1}$.
Equation (\ref{suppor}) then can be interpreted as imposing a
suppression of the radiation flux by more than 17 orders of magnitude!

The mathematics, and most of the issues of physical interpretation,
leading to eq.~(\ref{suppor}) are not controversial.  From the
beginning of the debate about the physical reality of the suppression,
the crucial question has been whether $\psi_0^0$ could be interpreted
as the quadrupole moment of the source, as is done in flat space.
What is really needed, of course, is a source calculation clearly
showing the relationship between $\psi_0^0$ and the source quadrupole
moment. The first approach to this was a scalar model given by
Kozameh, Newman and Rovelli \cite{KNR}.  Two of us (RP and JP) did an
explicit calculation \cite{PP1} for a compact source at the center of
a spherical, perfect fluid ``galaxy.''  A Green function solution gave
the relationship between the source and the waves outside the galaxy,
and showed no evidence for suppression. This calculation, however, was
not done directly in terms of $\psi_0^0$. It could not, therefore,
directly reveal the ``enhancement'' that must appear in a calculation
of $\psi_0^0$ in order to offset the mathematical suppression present
in eq.~(\ref{suppor}).  At about the same time, one of us (PK) wrote
down the form of the Green function solution directly in terms of
$\psi_0^0$, and found no evidence for this
enhancement \cite{K2,KPreprint}, suggesting that the
suppresion may be a real physical effect.

Here we resolve the apparently divergent findings.
To describe the unperturbed background spacetime, both inside and
outside the galaxy, we follow the notation of \cite{PP1} and take the
form of the metric to be
$ds^2=-e^\nu dt^2+e^\lambda dr^2+r^2(d\theta^2+\sin^2\theta d\phi^2)$
with $\nu$ and $\lambda$ functions of $r$ only.
We define the radial variable $r_*$  by
$dr/dr_*\equiv e^{(\nu-\lambda)/2}\equiv e^{\alpha(r)}$
and the retarded time $u$  by
$u\equiv t-r_*$.

We treat the source of gravitational waves as a perturbation on the
background of the metric given above, and we write
$\tilde{\Psi}_0=\hat{\Psi}_0(r,\omega) e^{i\omega t}\ \ \
\tilde{\psi}^0_0=\hat{\psi}^0_0(r,\omega) e^{i\omega t}$ for an an
$\ell$-pole moment with time dependence $e^{i\omega t}$.  The
perturbed field equations in general relate perturbations of the Weyl
projections, the NP spin coefficients, and the stress-energy
perturbations. For clarity of description, we will assume that the
equations can be combined to give a decoupled equation for $\hat{\Psi}_0$
of the form
\begin{equation}\label{vagueq}
{\cal D}\hat{\Psi}_0= \hat{\Psi}_0'' +f(r,\omega)
\hat{\Psi}_0'+g(r,\omega) \hat{\Psi}_0
=S(r,\omega)\,,
\end{equation}
in which the source function $S(r,\omega)$ is known, in which the
coefficient functions $f$ and $g$ are known functions constructed from
the background metric, and in which prime denotes differentiation with
respect to $r$.  There are at least two cases for which such an
equation can explicitly  be found: (i) if the ``galaxy'' is made of
perfect fluid and the waves are odd-parity, as in \cite{PP1},
and (ii) if the source lies in the vacuum exterior of a galaxy or hole,
as in \cite{K2,KPreprint}. The statements below about the radial dependence
of the Wronskians, and other functions, refer to calculations made in
these two cases.

We choose boundary conditions for eq.~(\ref{vagueq}) appropriate to
the galactic center and for outgoing waves far from the galaxy.  To
construct a Green function satisfying these conditions we define two
solutions of the homogeneous equation
${\cal D}{\cal R}=0$.
The function ${\cal R}_{\rm c}$ with the limit
${\cal R}_{\rm c}\stackrel{r\rightarrow0}{\longrightarrow} r^{\ell-2}$
has the correct behavior at the center of the galaxy, while
${\cal R}_{\rm w}$
with the limit
${\cal R}_{\rm w}\stackrel{r\rightarrow\infty}{\longrightarrow}
r^{-5}e^{-i\omega r_*}$
represents outgoing waves.

In terms of these functions it is straightforward to write the
solution to eq.~(\ref{vagueq}), outside the source, as
$\hat{\Psi}_0={\cal R}_{\rm w} \int\frac{S(\tilde{r},\omega){\cal
R}_{\rm c}}{W({\cal R}_{\rm c}, {\cal R}_{\rm w})}d\tilde{r}$ where
$W({\cal R}_{\rm c}, {\cal R}_{\rm w})$ is the Wronskian ${\cal
R}_{\rm c} {\cal R}'_{\rm w}-{\cal R}'_{\rm c}, {\cal R}_{\rm w}$.
{}From these definitions we arrive at a Green function solution
\begin{equation}\label{GF4psi}
\hat{\psi}^0_0=\int
\frac{S(\tilde{r},\omega){\cal R}_{\rm c}}{W({\cal R}_{\rm c},
{\cal R}_{\rm w})}d\tilde{r}
\end{equation}
for  $\psi^0_0$.

It will be useful in discussing this result for us to consider a case
(like that of a neutron star binary in an ordinary galaxy) for which
both the galaxy and the gravitational wave source are nonrelativistic,
and for which $\omega M$ is enormous. In this case, eq.~(\ref{suppor})
predicts enormous suppression.  If the suppression is a mathematical
artifact, not a physical suppression, we must find in
eq.~(\ref{GF4psi}) a counterbalancing enhancement factor.  Since the
galaxy is nonrelativistic, we can immediately eliminate several
possibile sources of such a factor.  The function ${\cal R}_{\rm c}$
cannot give rise to the enhancement factor since it retains the same
form as in flat space at the center.  The source term
$S(\tilde{r},\omega)$ is constructed from the source stress-energy
(which cannot contain a reference to $M$ outside the galaxy), and from
the spacetime geometry at the region of the source (which is only
influenced by a negligible $M/R$ contribution.

We conclude that the numerator in the integrand in eq.~(\ref{GF4psi})
is negligibly different from what it would be in a flat spacetime
background. In particular, it cannot contain a large enhancement
factor of the form $\left[1\pm 12\frac{(\ell-2)!}{(\ell+2)!}iM\omega
\right]$. If such an enhancement factor is to appear it must come from
the Wronskian, and the Wronskian, unlike the other terms in
eq.~(\ref{GF4psi}), cannot be ruled out as the source of such a
factor.  The Wronskian contains solutions normalized both at the
center of the galaxy and in the exterior, and hence ``knows'' what the
mass of the galaxy is. Aside from small corrections (such as the central
redshift), of order $M/R$, we find that at $r\rightarrow0$
\begin{equation}\label{Wcenter}
W({\cal R}_{\rm c},{\cal R}_{\rm w})=\kappa_1\left[1
-12\frac{(\ell-2)!}{(\ell+2)!}iM\omega
\right]^{-1}\,,
\end{equation}
where $\kappa_1$ is a function of $\omega,r$ and $\ell$ which has the
same value it would have in flat spacetime.

We conclude that  eq.~(\ref{GF4psi}), for a
compact source at the center of a nonrelativistic galaxy, reduces to
\begin{equation}\label{Kenh}
\hat{\psi^0_0}= \kappa_2 \left[1
-12\frac{(\ell-2)!}{(\ell+2)!}iM\omega
\right]\,,
\end{equation}
where $\kappa_2$ has the same value it would have in a flat background
except for (negligible) corrections of order $M/R$. The enhancement
factor, in the square brackets in eq.~(\ref{Kenh}), cancels the
suppression factor in eq.~(\ref{suppor}), and we conclude that in this
situation $\hat{\sigma}_0$ is the same (aside from negligible
corrections of order $M/R$) as it would be in flat spacetime. {\em
Aside from the small redshift effect, the gravitational field of the
galaxy has no consequences for the emergence of radiation produced at
its center.}

The above argument applies only for a source at a distance from the
galactic center small compared to the wavelength of the radiation it
produces. As the source location [$\tilde{r}$ in (\ref{GF4psi})] moves
outward, there are significant changes in the forms both of $W({\cal
R}_{\rm c},{\cal R}_{\rm w})$ and of $S(\tilde{r},\omega)$.  We can
see the trends most clearly if we consider the source to be in the
Schwarzschild exterior of the galaxy. In the exterior (\ref{Wcenter})
is replaced by $W({\cal R}_{\rm c},{\cal R}_{\rm w})=\kappa_1$ where
corrections of order $M/R$ have been omitted.  The enhancement factor
needed to cancel the suppression in (\ref{suppor}) is now missing.
It should be emphasized that this change in the character of the
Wronskian is the key to resolving previous apparently contradictory
results.  The nontrivial behavior, in (\ref{Wcenter}), of the
Wronskian near $r=0$ was not anticipated when the original arguments
for suppression were made.

More than the Wronskian changes when the source is moved to the exterior;
there is also an important change in ${\cal R}_{\rm c}$.
In the Schwarzschild exterior, aside from small
corrections, ${\cal R}_{\rm c}$ takes the form
\begin{equation}\label{Rcoutsimple}
{\cal R}_{\rm c}=\kappa_3e^{i\omega r_*}+\kappa_4
\left[1
-12\frac{(\ell-2)!}{(\ell+2)!}iM\omega
\right] e^{-i\omega r_*}\,,
\end{equation}
in which the $\kappa_i$ have the same form as for a flat background.
In the exterior,
therefore, the form of ${\cal R}_{\rm c}$ supplies an enhancement
factor for {\em part} of the source integral in (\ref{GF4psi}). For
this part of $\hat{\psi}^0_0$ the enhancement factor will cancel the
suppression factor in (\ref{suppor}) and the contribution to the
radiation will be the same (aside from small corrections) as in a flat
background. But the remainder of the source integral for
$\hat{\psi}^0_0$, that due to the $\kappa_3$ term in
(\ref{Rcoutsimple}), will be reduced by the suppression factor.

The situation is somewhat similar for sources in the vicinity of a black
hole, but some details must be altered. In the case of galaxies, we
used the homogeneous function ${\cal R}_{\rm c}$, normalized at $r=0$.
For holes we use instead the function ${\cal R}_{\rm hole}$
representing waves ingoing at the horizon.
The Wronskian in the Green function then has the value
\begin{equation}
W({\cal R}_{\rm hole},{\cal R}_{\rm
w})=\frac{1}{T_\ell}\,\frac{K(\ell)}{\omega}\,
\frac{8i\omega^3}{r^6(1-2M/r)^3}\,,
\end{equation}
and at large $r$ the form of ${\cal R}_{\rm hole}$
is \begin{equation}\label{holeout}
{\cal R}_{\rm hole}\stackrel{r\rightarrow\infty}{\longrightarrow}
\frac{\kappa_5}{T_\ell(\omega)}\left\{
\kappa_3e^{i\omega r_*}+\kappa_4 R_\ell(\omega)
\left[1
-12\frac{(\ell-2)!}{(\ell+2)!}iM\omega
\right] e^{-i\omega r_*}
\right\}\,.
\end{equation}
Here $\kappa_3$ and $\kappa_4$ are the same as in (\ref{Rcoutsimple}),
and
$K(\ell)$
and $\kappa_5$ have the same value as in flat spacetime.  The factors
$T_\ell(\omega)$ and $R_\ell(\omega)$ are the transmission and
reflection coefficients for gravitational waves.  For high
frequencies ($\omega M\gg1$) the transmission coefficient is
negligibly different from unity, and the reflection coefficient is
negligibly small, at least for quadrupole and other low $\ell$-pole
moments.
Since the enhancement factor is only present for the $\kappa_4$ term
in (\ref{holeout}) it is the only part that will not be reduced --
relative to the flat space value -- by the suppression factor in
(\ref{suppor}). Since the reflection coefficient should be small, the
result gives the appearance of significant suppression of
radiation for a compact source outside a hole.

The first step in trying to understand these results is an often
overlooked point about sources: the emission from a compact
``quadrupole'' (i.e., nonrelativistic) source, located far from the
coordinate center, {\em will not be dominated by $\ell=2$.} A compact
source, far from the coordinate origin, radiates predominately high
$\ell$ multipoles. {\em A distant source, even a nonrelativistic
``quadrupole'' source, radiates a negligible fraction of its power at
low $\ell$-pole moments.}

Source integrals combine the source stress-energy and a solution of
the homogeneous wave equation.  Near $r=0$ the homogeneous solution
has a power-law form, and the source integrals {\em if confined to the
region near $r=0$}, take the form of integrals for the various
multipole moments of the source mass distribution.  For a source
located at large $r$ (i.e., not within a small fraction of a
wavelength of $r=0$) the result is very different. In this case the
source integral for $\ell$-pole radiation is not related to the
$\ell$th multipole moment of the mass distribution of the source.
Rather, the radiation will be dominated by the multipoles which couple
best to the source distribution.  For a nonrelativistic source, of
wavelength $\lambda$ at radius (Schwarzschild radial coordinate value)
$R_{\rm source}$, the best coupling will be for $\ell\sim\ R_{\rm
source}/\lambda$.

We must therefore consider large values of $\ell$ when we consider
the suppression factor
\begin{displaymath}
\left[1
-12\frac{(\ell-2)!}{(\ell+2)!}iM\omega
\right]\approx1-12i M\omega/\ell^4\ .
\end{displaymath}
Typical numbers for a compact source and a galaxy are
$\omega\sim10^{-7}{\rm cm}^{-1}$, $R\approx10^{23}$cm, $M\approx
10^{16}$cm. The distance to an exterior source must be at least as
large as the galaxy radius, and this means that the radiation will
characteristically be at $\ell\sim\ R_{\rm source}/\lambda\geq
R/\lambda$, and $R/\lambda$ is on the order $10^{15}$.  The
suppression factor then differs from unity by a correction
$12M\omega/\ell^4$ of order $10^{-37}$ or smaller.

Thus for ``typical'' sources and galaxies, the suppression factor does
not play a significant role in determining the radiation reaching a
distant observer. Can one think of sources ---at least in principle---
for which the mathematics indicates that there is significant
suppression, but intuition demands that there is not? We can easily
argue that no such situation can arise.  First, for suppression to be
important $M\omega$ must be large.  Second, the distance to the source
$R_{\rm source}$ must be no smaller than the Schwarzschild radius $2M$
of the galaxy. Thus we have
\begin{equation}
\ell\sim \frac{R_{\rm source}}{\lambda}\geq \frac{M}{\lambda}\sim
M\omega\ .
\end{equation}
It follows that $M\omega/\ell^4$ is no larger than order
$(M\omega)^{-3}$ and hence cannot be large.  The suppression factor
can have an important effect only for a source just outside a black
hole ($R_{\rm source}\sim M$), with source wavelength on the order of
the radius of the hole ($M\omega\sim1$). But in these circumstances we
would certainly expect the curved background to influence the
emergence of radiation! One example is a test particle falling
radially into a Schwarzschild black hole of mass M. The early work of
Davis {\em et al.} \cite{DaRuPrPr} shows that the emitted radiation is
predominantly quadrupolar and that the spectrum peaks at
$\omega=\omega_{max}\approx 0.32M^{-1}$. At the peak frequency the
suppression factor for $\ell=2$ waves is $1+0.16i$, yielding a small
but nonnegligible reduction (of about 3\%) of the radiated energy per
unit frequency interval $dE/d\omega$. This reduction, relative to the
calculated radiation for a flat background, is presumably already
included in the numerical results, since the calculation took explicit
account of the strongly curved background.

We can then conclude that {\em for any configuration of source and
galaxy or black hole, the suppression factor can have a significant
effect only in the case ($R_{\rm source}\sim M\sim\lambda$) that a
significant effect would be predicted on the basis of the standard
viewpoint.}

We thank Ted Newman,
Peter Saulson, and Ranjeet Tate for useful discussions.  Two of us (RP
and JP) gratefully acknowledge the support of grant NSF PHY92-07225.

\end{document}